# Observing dynamos in cool stars

Zs. Kővári & K. Oláh

*Konkoly Observatory of the Hungarian Academy of Sciences, Konkoly Thege út 15-17., H-1121, Budapest, Hungary*
zsolt.kovari@csfk.mta.hu, katalin.olah@csfk.mta.hu

**Abstract** The main aim of this paper is to introduce the most important observables that help us to investigate stellar dynamos and compare those to the modeling results. We give an overview of the available observational methods and data processing techniques that are suitable for such purposes, with touching upon examples of inadequate interpretations as well. Stellar observations are compared to the solar data in such a way, which ensures that the measurements are comparable in dimension, wavelength, and timescale. A brief outlook is given to the future plans and possibilities. A thorough review of this topic was published nearly a decade ago (Berdyugina 2005), now we focus on the experience that have been gathered since that time.

## 1 Introduction

For now it goes without saying, that, likewise sunspots, starspots are fingerprints of stellar magnetic fields. On the other hand, magnetic fields induce local and global brightness variability either on shorter or on longer terms. Therefore, observing starspots on cool stars is unavoidably important to get constraints for solar and stellar dynamo theory. We have learned that to a certain extent, solar magnetic activity can serve as a proxy of active stars of different types. However, we have already learned as well, that often the stellar observations cannot be understood within the solar dynamo theory. Therefore, it is also necessary to review the most important parameters, features or processes of the solar dynamo, of which stellar counterparts can be deduced from observing active stars (cf. Strassmeier 2005, 2009). Such experiences give base for studying how the observable phenomena or quantities are compatible with the solar and stellar dynamos and which parts of the theory require revision (cf. Kitchatinov & Rüdiger 2004).

So, what are the dynamo ingredients in cool stars that can be observed? A short list of the most important ones would include: (i) cycles of different types observed at different wavelengths; (ii) rotational period variations related to e.g., surface spot activity or differential rotation (iii) preferred spot locations, active longitudes; (iv) flip-flop phenomenon, flip-flop cycle; (v) local and global magnetic fields; (vi) local and global surface flows.

Note, that some of the listed stellar dynamo observables require long-term data acquisition, while others need the most advanced technologies and the most modern observing facilities. Individual objects can be studied with 0.5–2.0-m class robotic telescopes which by now fully replaced the manpower in systematic long-term data gathering. The continuous monitoring of a number of active stars have basically two useful timescales: the order of the rotation period and a much longer one covering the activity cycle(s). Long-term continuous photometry on the timescale of the activity cycles provides information on the overall brightness variability, i.e., on the overall spot coverage, therefore serving as comparison and constraints to various dynamo models. On the other hand, high-resolution spectroscopy on the timescale of a few rotations could provide us information on the surface differential rotation and also on other global and local surface flows. However, this demands the use of 2–4-m class telescopes.

In the recent years, ultra-high precision space-photometry has opened the perspective of statistical study of activity, but also has provoked the development of new data processing techniques as a consequence.

However, beside the precision, an important factor of studying active stars is time. Observing the long-term behavior of active stars, i.e., cycles, needs continuous monitoring for decades, which is not a fashionable program and does not need high-tech developments. On the contrary, the maintenance of the instruments as long as possible (and if necessary, a very careful replacement) is a key factor in getting homogeneous datasets. Only long and reliable datasets can constrain the dynamo models, which describe theoretically the magnetic behavior of the Sun and stars. The small, 1-m class telescopes have a key role in this problem, since the brighter stars are the ones which already have long records of observational data and the extension of these series is the most useful for studying stellar cycles.

## 2 Solar and stellar cycles

The Sun, because of its proximity, is an ideal target for studying magnetic cycles on different timescales and on different wavelengths. However, the adaptation of the knowledge deduced form the detailed and long-term solar observations to active stars has many problems and even could be misleading. The reasons are as follows: (i) magnetic activity are observed on very different kinds of stars concerning spectral type, mass and age, (ii) the lengths of the stellar datasets are much shorter than that of the Sun and do not exceed a few decades, and (iii) the datasets in most cases are different concerning frequency of the observations (time-base, regularity), the wavelength ranges, and the methods the data are gathered. In this section we compare the available information on the solar and stellar cycles, based on similar datasets.

### *2.1 Solar cycles from 1-D data - observations*

Sunspots are known for thousands of years, and the fact that their number on the solar surface is changing has been also recorded for a few hundred years already. This valuable dataset, and its extension in time by proxies are among the most important observational backgrounds in studying the solar dynamo. Unfortunately, stellar surfaces cannot be spatially resolved, and thus, deriving a quantity for spotted stars such as the Wolf-number for the Sun is not possible. Instead, fractional spot coverage (in percent of the total stellar surface) of some stars are followed for 1–2 decades at most - but no comparable measure in a long-term base exists on the Sun as to spot coverage. Sunspot numbers are thoroughly studied in other papers of this proceedings. We do not consider sunspot number showing the well-known cycles useful for a comparison to stellar cycles, except the nominal values of the cycle lengths.

Fortunately, long-term datasets exist which were obtained as measurements taken on the Sun as if it was a star. The most important and extremely valuable records are the 10.7 cm radio measurements taken each day for more than half a century. Similarly, the Geosynchronous Operational Environmental Satellite (GOES) provides X-ray flux of the Sun, continuously since 1994.

In one way we can use the measurements of the total solar irradiance, though, similar long-term data do not exist for stars. The spectral solar irradiance (Shapiro et al. 2011), which uses the minimum flux of the Sun combined with the sunspot numbers and proxies of the sunspot number, gives one average spectrum of the Sun each year for the past 400 years, between 130nm–1000nm. For the earlier time, one average spectrum for every 22 years, i.e., one for every magnetic cycle, is calculated. The wavelength-flux tables allow to get similar magnitudes of the Sun than those used for stellar measurements. This way we are able to seek for cycles of the Sun e.g., in the usual Johnson *B* and *V* bandpasses established for stellar observations, and from the reconstruction any other special flux values can easily be calculated.

## *2.2 Stellar cycles - observations*

The most important factor in studying activity cycles is the length of the available data. For the Sun, its activity is recorded for hundreds of years and this database is extended to a much longer timescale (millennia) using different proxies recorded on Earth which reflect the changes of the solar activity. In this review, for comparison purposes, we restrict the solar dataset to about a century, which is comparable to the longest available datasets for stars, using digitized plate collections (Digital Access to a Sky Century @ Harvard - DASCH). In Fig. 1 we show the lengths of the stellar cycles compared to the solar ones, and mark the maximum lengths of the available datasets from photometric monitoring and from the plate collections. It is evident, that, if we had data of the solar activity only as long as the longest stellar datasets (about 110 years), then, we could just guess that a longer solar cycle (i.e., the Gleissberg cycle) exists, but nothing would be known about the even longer timescale solar variation.

The first systematic search for cycles on stars which were suspected to have chromospheric activity observed as emission in their Ca II H&K lines, started in the late 1960's by Olin C. Wilson (1968), and the effort, which lasted for several decades ceased, unfortunately, in 2003. The so-called Mt. Wilson database produced the first insight to stellar activity cycles and the measurements of that project are being used to date. Recording starspots through photometric data started in the mid 1960's, first just sporadic data have been gathered on a few interesting objects. The breakthrough was the advent of the Automated Photoeletric Telescopes (APTs), which were specifically designed for following starspot activity on a long-term base. By now two such projects keep on getting multicolor data already for decades, the two Vienna-Potsdam APTs and the Four College Consortium APT. The lengths of the datasets from the Vienna-Potsdam APTs by now exceed 25 years. Observations taken in different bandpasses allow to estimate the average spot temperatures. Therefore, an important factor of using APT data for studying long-term behavior of stellar activity is the homogeneity of the data, which cannot easily be achieved by putting together measurements from different instruments.

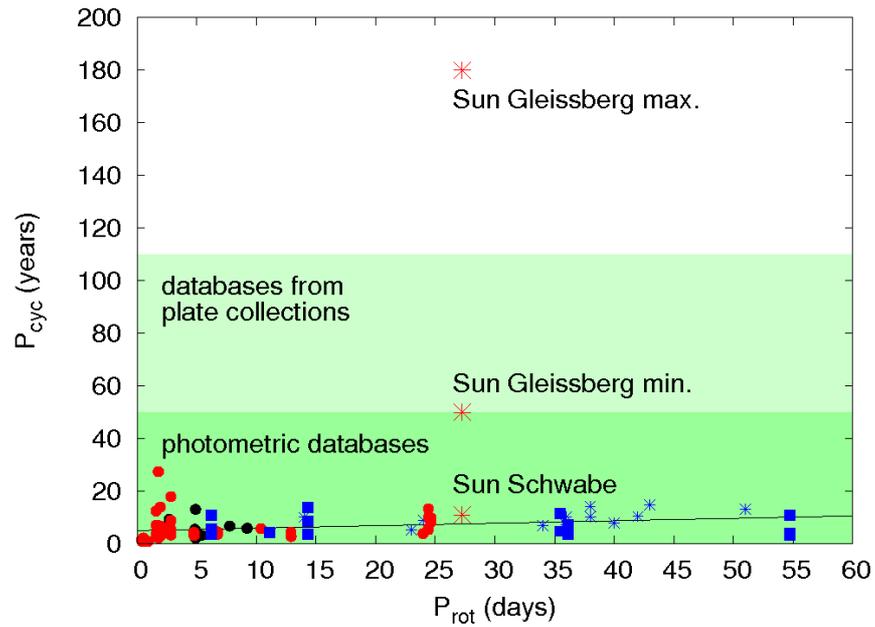

*Fig. 1: The known cycle lengths of active stars in the function of their rotational periods, the solar Schwabe cycle, and the shortest and longest known values of the Gleissberg cycle. Dark shaded area at the bottom bounds the length of the available photometric databases, while the lighter area above marks the limits of the available photographic databases. Different symbols represent cycles from different publications. A rough relation between the rotational periods and all activity cycles of stars is also indicated. The cycle periods are from Oláh et al. (2009) and references therein.*

Other presently running surveys, implemented not specifically for stellar activity research produce, however, useful information on the long-term behavior of active stars as well, now for more than a decade. One of these programs is the All Sky Automated Survey (ASAS, Pojmanski, 2002), which monitors stars brighter than 14 magnitude all over the sky. Its public database provides systematic observations in *V*-band for many active stars. At present, the ASAS telescopes (both in the northern and southern hemispheres) are gathering data simultaneously in *V* and *I* bandpasses. The HATNet survey telescopes have originally been designed for searching extrasolar planets (Bakos et al. 2002), but also resulted in a large number of observations of active stars, and from that such statistical results followed like rotation-age-activity-mass relation (Hartmann et al. 2011). Details of the various automated telescopes used in stellar activity are found in Berdyugina (2005, in Sect. 3)

The next step, about a decade ago, was the introduction of the automated telescopes constructed for collecting spectral observations of active stars on a long-term base, in order to get Doppler images (see Sect. 3.1.2 for details of Doppler imaging) of the stars regularly, and this way to find cycles in the spot coverage. Precise $v \sin i$ measurements are also gathered for orbital solutions with active stars in binary systems. The TSU 2-m Automatic Spectroscopic Telescope in Arizona started observing in 2004 while the 1.2-m STELLA-I in Tenerife in 2006. Since then, a number of publications have been based on data by these robotic facilities from the first decade of their operations, see, e.g., Fekel & Bolton (2007), Korhonen et al. (2009), Strassmeier et al. (2010, 2011, 2012), Fekel et al. (2013), etc.

However, in more extreme wavelengths like X-ray or radio, no systematic project exists for observing stellar activity, except the Sun. Still, long-term behavior of a few active stars can be studied by combining data from different space missions like *Einstein*, *ROSAT* and *XMM Newton*, in X-ray.

Recently, there is a new possibility to get stellar cycles through the systematic variation of the rotational periods due to differential rotation, similar to that observed on the Sun as butterfly diagram. The ultra-high precision and continuous datasets observed by the K1 mission of *Kepler* make this type of investigation possible. The drawback is the relatively short time-base though, i.e., only stars with very rapid rotation (in the order of 0.5 day) could be studied, since for these objects the shortest cycle is suspected to be in the order of 1 year.

Magnetic activity and its variability have an impact on pulsation frequencies of the Sun and stars. Promising results in the solar case have been published by Jiménez et al. (2011), showing clear correlation of the acoustic cutoff frequency and the solar activity cycle. Since the solar cycle is well followed, the observed relation is on firm grounds. Regarding the stars, from CoRoT data of the solar-like

HD 49933 García et al. (2010) found correlation between the oscillation frequencies and mode amplitudes, and the luminosity changes due to magnetic activity.

## *2.3 Solar and stellar cycles - results and comparisons*

When speaking about activity cycles one should keep in mind that the activity phenomena are changing in time but not absolutely regularly. This means that the cycle lengths are not periods but rather timescales, i.e., in the datasets no strict periods are found. Similarly, due to the differential rotation, which acts in most types of active stars with various strengths, the stellar rotational periods derived from photometry can be different on the same star, depending on the latitudes populated by the star spots. Thus, in case of active stars, the conventional methods for period search designed for strictly periodic signals should be applied with caution.

Throughout this paper, solar activity cycle, in short, solar cycle, is considered with a length of about 11 years. This triviality is mentioned because solar cycle is also understood as magnetic cycle with double length. But at present, there is no observational evidence of stellar magnetic cycles from magnetic field measurements, thus only cycles from photometry, fluxes in other wavelengths, and spectral indices of the Sun and stars, are comparable.

### 2.3.1 Solar-type stars

The 11-year Schwabe cycle of the Sun is well followed in various wavelengths. Fig. 2 shows the variability of the solar X-ray flux during the last 20 years. Apart from the strong short-term variability due to the rotation and flare activity, the 11-year long cycle in the mean values is axiomatical.

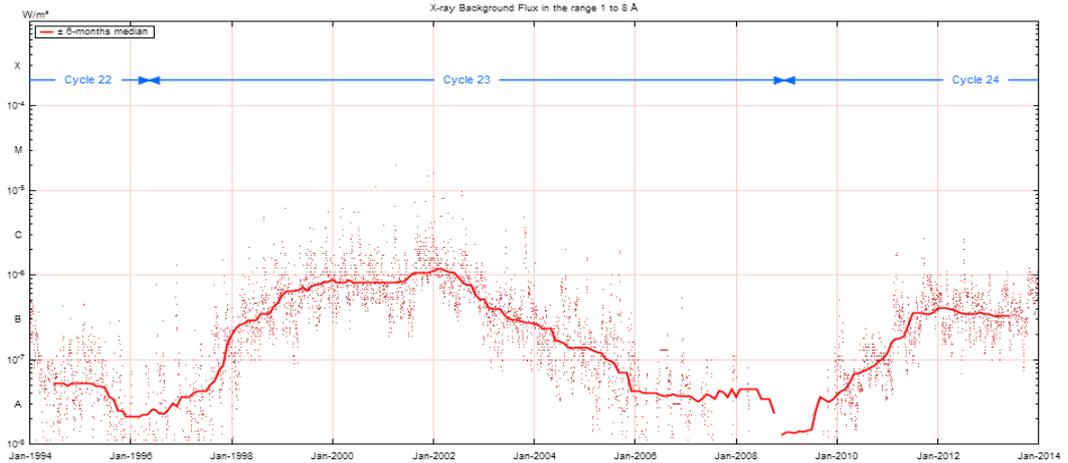

*Fig. 2: Solar X-ray flux between 1-8Å from GOES data (http://sidstation.loudet.org/solar-activity-en.xhtml)*

AB Dor is a nearby (*d*=15 pc) bright (*V*=6.75) quadruple system, of which the brightest component AB Dor A is an active K0-dwarf. It is an ideal target of observations in many wavelengths; more than a decade long data are available in the 0.3-2.5 keV X-ray bandpass (Lalitha and Schmitt 2013). In Fig. 3 the X-ray behaviour of AB Dor A is seen between 2000-2010, showing large fluctuations due to flares and a marginal long-term variation, still resembling of cyclic behaviour.

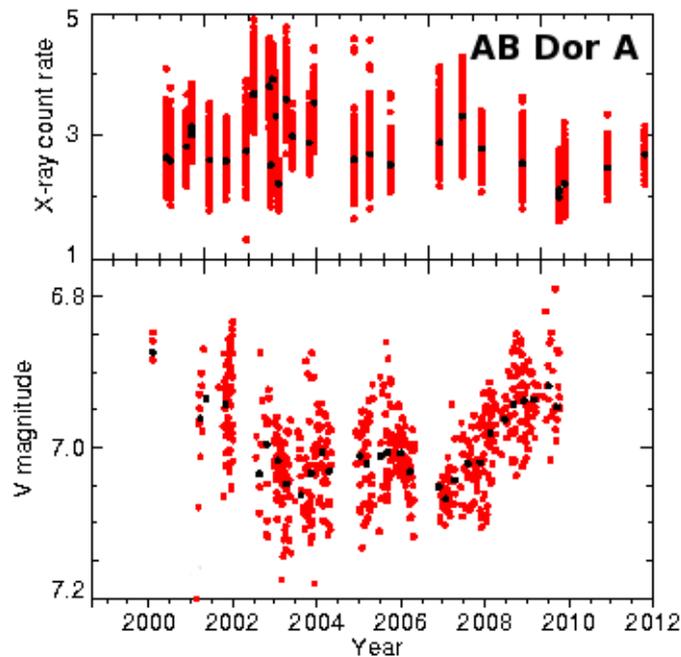

*Fig. 3: AB Dor A X-ray luminosity between 0.3-2.5 keV (top) together with V magnitude (bottom) from Lalitha and Schmitt (2013). A marginal long-term change in the X-ray count rate is seen, resembling to a cyclic behaviour.*

The solar X-ray luminosity changes between log $L_X$=26.8 and log $L_X$=27.9 during a solar cycle in the *ROSAT* bandpass of 0.1-2.4 keV, between cycle minimum and maximum (Judge et al., 2003), i.e., it changes one order of a magnitude. AB Dor A shows changes between log $L_X$=29.8-30.2 in a similar bandpass of 0.3-2.5 keV, which is on average two magnitudes stronger, but has a much lower amplitude than that of the Sun.

For the Sun it is widely accepted that it becomes bluer when more active, i.e., at spot maximum the Sun radiates more than at minimum, which is obvious from the irradiance measurements. However, no long-term irradiance measurements exist on any active star, therefore no direct comparison is possible. In Shapiro et al. (2011) yearly mean values are deduced from the spectral reconstruction in *B* and *V* bandpasses (broadband *B*-filter is at 445 nm, *V*-filter is at 551 nm effective wavelengths, with FWHM of about 90nm), and from these one can get similar brightness-color index diagrams for the Sun as those of Messina (2008) for stars. From the reconstructed *B* and *V* colors it is found, that the Sun becomes redder when fainter, although with very small amplitude (Oláh et al. 2012). This unsuspected result, however, is in accordance with Preminger et al.'s (2011) finding that the total solar *visible* continuum brightness is decreasing, when the solar activity is increasing.

Activity cycles on solar-type stars from Ca-index measurements were published by Baliunas et al. (1996), first time describing a relation between the rotational and cycle periods, and connecting this relation to the stellar dynamo in the seminal paper *"A Dynamo Interpretation of Stellar Activity Cycles"*. The dataset describes the variation of the stellar chromospheres. Using contemporaneous Ca-index and photometric data Radick et al. (1998) found a strong connection between the behavior of the photospheres and chromospheres of active stars, showing direct or anticorrelation, thereby differentiating between the spot- and plage-dominated activity. Messina (2008) gave long-term *UBV* photometry for 14 well-known active stars studying their color index behavior as a function of brightness. The results revealed that part of the stars are spot-dominated while the rest show photospheric faculae dominance.

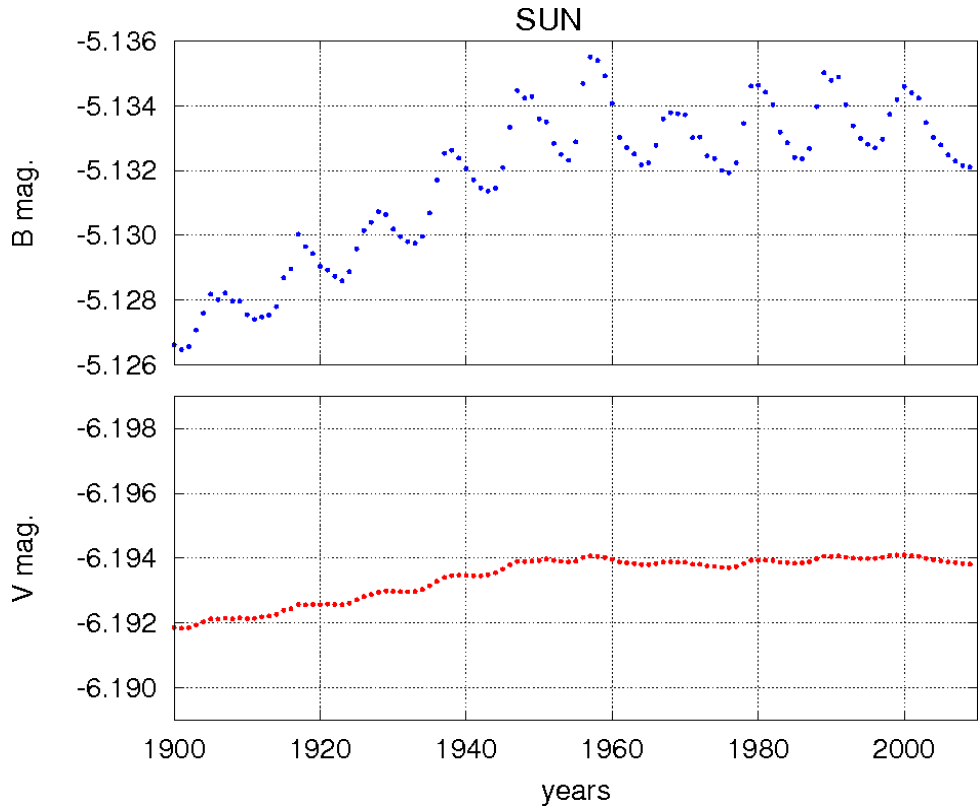

*Fig. 4: The brightness of the Sun in Johnson B and V bands, for about a century, based on Shapiro et al.'s (2011) reconstruction. A long-term change is well seen (Gleissberg cycle), the 11-yr cycle is evident in B color with a small amplitude, and marginal in V. Magnitude scales are arbitrary without zero point.*

Fig. 4 shows the cyclic behavior of the Sun in *B* and *V* colors (Shapiro et al. 2011), for the last century only. As seen from Fig. 4., in *B* color the Schwabe cycle is evident and a longer-term change (part of a Gleissberg cycle) is also seen, though the amplitude is small. In *V* color the amplitude is even smaller, the 11-year cycle is barely seen, but the longer term variation is clear. Note that the Sun shows a very marginal light variation in *V* color, about 0.002 mag in the course of its cycles.

BE Cet is a single, relatively young (cca. 600 Myr, member of the Hyades moving group) solar-type star (G2V class) with a rotational period of 7.76 days, i.e., a solar analog in the sense the Sun in its youth could have been like this active star. BE Cet shows rotational modulation due to starspots and long-term cyclic change of about 6.7 years with a low amplitude of about 0.02 mag. (see Fig. 5 and Messina & Guinan 2002) which is still an order higher than the solar amplitude.

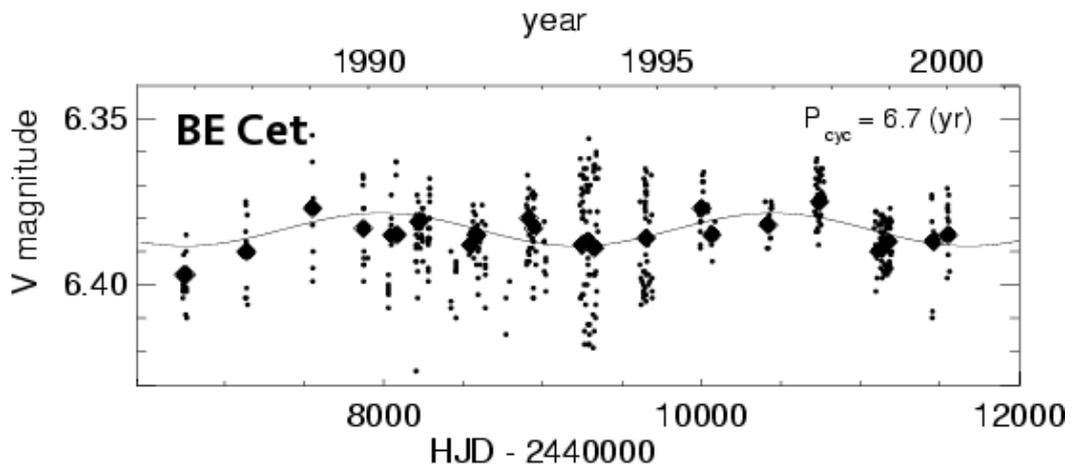

*Fig. 5: Long-term variability of the the solar-like star BE Cet. Sinusoidal fit to the data suggests a cycle length of 6.7 years (Messina and Guinan, 2002).*

An even younger (about 10-30 Myr old only) solar analog star is EK Dra (G1.5V), an effectively single dwarf (its M-dwarf companion has an orbital period of 45 years), which has a rotational period of 2.6 days, i.e., rotating 10 times faster than the Sun. Its rotational modulation as well as cycles have much higher amplitudes compared to BE Cet, amounting to a few tenths of magnitudes according to Järvinen et al. (2007); see also Fig. 6.

The rotational rate plays a basic role in the strength of the magnetic activity. Amplitudes of the cyclic changes are much higher for stars rotating faster. (However, one should not forget that the *observed* amplitude of the rotational modulation depends on the inclination angle as well.) The different rotational modulation and cycle amplitudes of the very similar spectral type stars BE Cet, EK Dra and the Sun reflect their rotational rates and ages, since the rotational rate decreases due to magnetic braking as the star evolves.

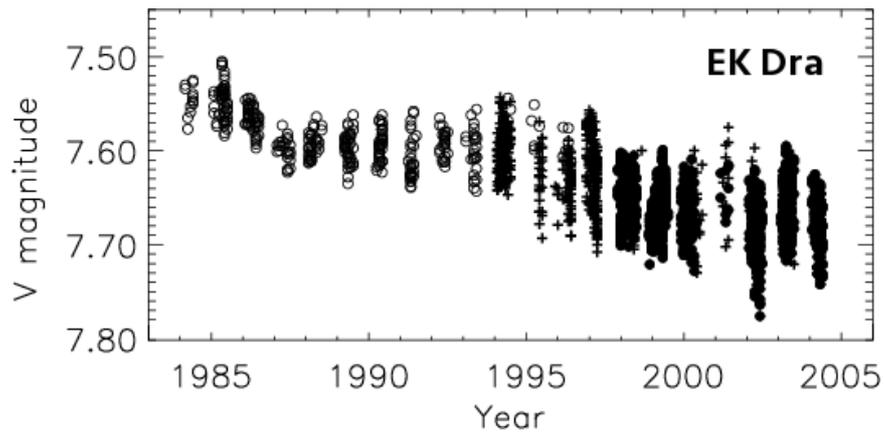

*Fig. 6: Long-term brightness change of EK Dra (Järvinen et al. 2007).*

## 2.3.2 Stars of different spectral types

Activity cycles are observed on many different kinds of stars: singles and binaries, dwarfs and giants, from F to M-type stars. Do we see differences between the cycles of these stars? How the cycles behave in time? It is known for a long time, that the length of solar cycles are variable, e.g., the Schwabe cycle varies between about 9-14 years and the Gleissberg cycle also varies continuously (Kolláth & Oláh 2009). To study the temporal behavior of the stellar cycles we need as long datasets as possible, but unfortunately, at present even a century long dataset of an active star is rare. Different types of time-frequency analysis tools are used to study the temporal variation of the activity cycles. However, time-frequency analysis requires continuous and equidistant datasets. The stellar long-term photometric records are biased (i) by the visibility of the target from a given location, (ii) by the effects of the weather conditions, and finally, (iii) by the distortion of the rotational modulation in case of long periods and not enough densely sampled observations. All these are the features of the ground-based observations. On the other hand, space-born data are more precise, can be continuous, equidistant and are barely subject of visibility. But their available time-base is still short compared to ground-based datasets, therefore only the shortest cycles of short period stars can be investigated from them.

In Kolláth & Oláh (2009) the time-frequency method of studying ill-sampled datasets is thoroughly discussed and a recipe was given how to overcome the problems. Briefly: first the rotational modulation is removed from the data, after that an appropriate spline interpolation is applied to get rid of the yearly gaps. The result of this procedure is a dataset for which time-frequency methods can be applied.

Based on the method given by Kolláth & Oláh (2009), in Fig. 7 we now can compare the long-term photometric variability of HK Lac, a K0-giant active primary of a close binary system, with a rotational (and orbital) period of 24.5 days, and the cyclic behavior of LQ Hya, a single active K2-dwarf, rotating with 1.6 days. The light variation of the two objects look very similar, decades-long variations are seen, as well as modulations on the timescale of a few years. Together with the previous examples of the Sun, solar analogs and different types of active stars, we find that the activity cycles are generally *variable* (no strict periods, only timescales can be determined), and *multiple*. In case of active stars the lengths of the databases are still not long enough to determine the longer cycle lengths in most cases.

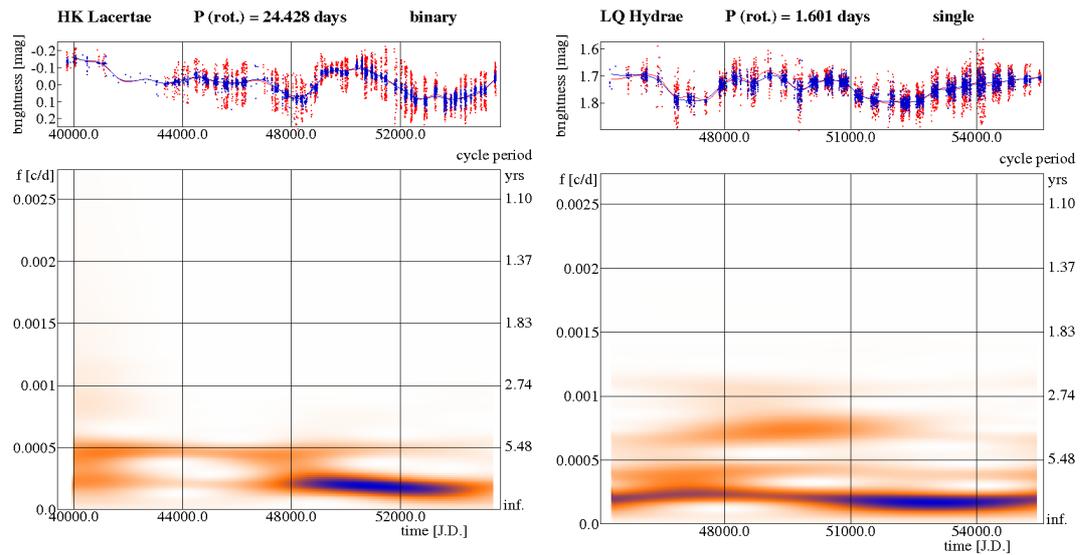

*Fig. 7: Time-frequency diagram from short-term Fourier-transform for the K0-giant HK Lac in a binary (left), and LQ Hya, a fast rotating single K2-dwarf (right). Based on Oláh et al. (2012).*

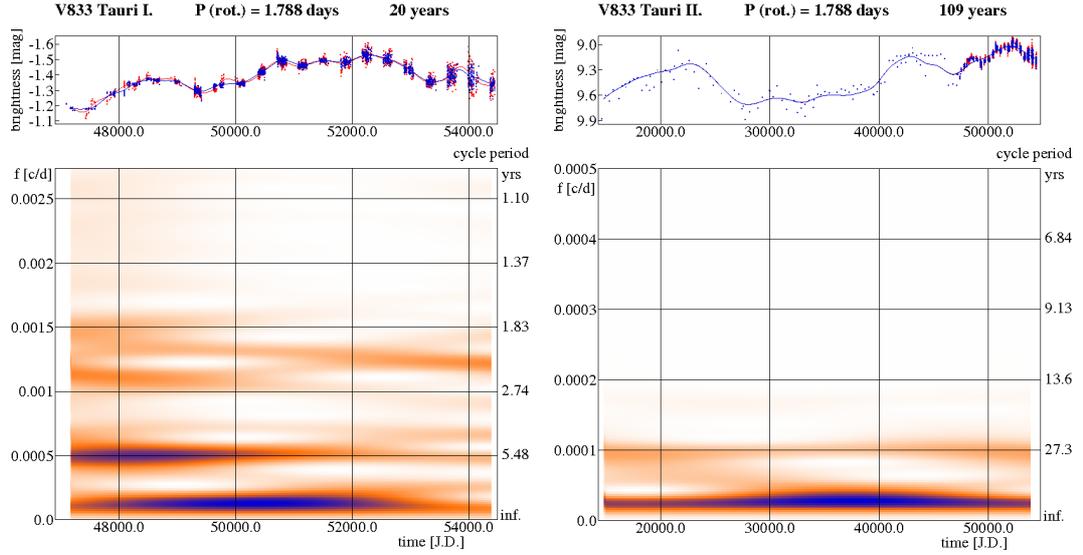

*Fig. 8: Cycles for V833 Tau: short-term Fourier-transform was applied for the 20 years long photometric observations (left), and for a larger time base extended with photographic data (right), based on Oláh et al. (2009).*

Luckily, for a few objects, putting together the photographic and photoelectric datasets we could have longer time base. One of such examples is V833 Tau, a K2-dwarf in a close binary system with a possibly brown dwarf companion (Cuntz et al. 2000). In the left panel of Fig. 8 the cyclic behavior of the primary star is seen from a 20 years long photometric dataset. The right panel shows almost a century-long photographic dataset plus the last 20 years of photometry, and the derived long-term changes. The photographic data are sparse, but they clearly show a much higher amplitude light variation than that during the 20 years of photometry. Because of the low inclination (≈20°) of the system, the rotational modulation due to spots have small amplitude and does not distort the sparse long-term data significantly when observing the star at different phases of the rotation. The 2-3, 5, 27-30 years and even longer term variations show the multicyclic nature of the brightness variability, which, on shorter timescales though, resembles of the solar cycles with 11, 60-120 and longer timescales.

## 2.3.3 Uncovering stellar butterfly diagrams

Solar butterfly diagram is one of the most striking examples of cyclic behavior of solar activity. During the 11-year sunspot cycle, the activity wave migrates from mid-latitudes to the equator, giving scope for observing the photospheric manifestation of the dynamo action underneath. The limited spatial resolution of stellar observations makes more difficult to compile such diagrams for spotted stars. In long-term photometric observations of a spotted star, a latitudinally migrating activity belt *together* with surface differential rotation would result in a changing peak-to-peak amplitude of the light curves. Katsova et al. (2003) and Livshits et al. (2003) suggested a model to construct stellar butterfly diagram from long-term photometric data, however, with rigorous *a priori* assumptions on the spot distribution. On the other hand, once the differential rotation is known (e.g., from Doppler imaging), results from light curve inversions for an extended period can be suitable to recover stellar butterfly diagram without involving any assumption on spot geometry (Berdyugina & Henry 2007).

Within the four years of the original *Kepler* mission (K1) ended in May 2013, a number of ultrafast-rotating active K-M-dwarfs were monitored. For such stars with $P_{rot}$ of ≈0.5 days Vida et al. (2013) found cycle lengths in the order of 400 days, i.e., already within the reach of *Kepler*; (cf. Fig. 10). According to the model calculations for a K0-dwarf with a rotational period of 2 days (Işik et al. 2011) the expected butterfly diagram will dramatically change, compared to the solar case, due to the high latitude spottedness and the much thinner (≈10°) activity belt. Still, with enough strong surface differential rotation, a small modulation of the rotational period will occur during the activity cycle. The typical amplitude of such a modulation is expectedly large enough to be detected from high-precision *Kepler* photometry. Vida et al. (2014) found this kind of behavior in one-fourth of their sample of 39 single fast-rotating active dwarfs (see Fig. 9).

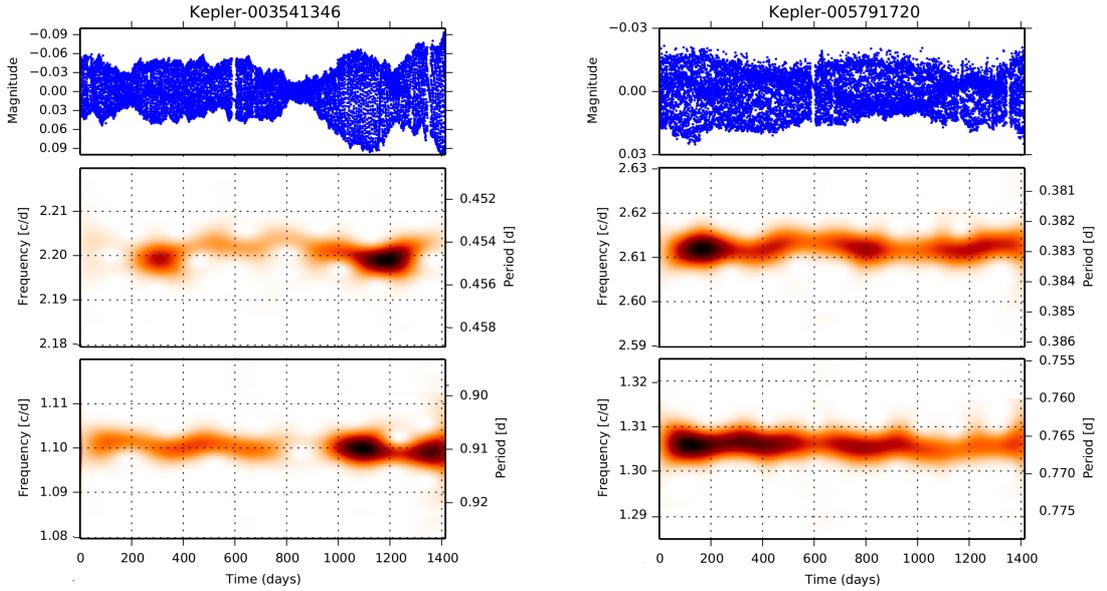

*Fig, 9: Two examples from Vida et al. (2014) for quasi-periodic modulation in the rotation period of fast-rotating active dwarfs (left: KIC 03541346, $P_{rot}$= 0.9082 days, right: KIC 05791720, $P_{rot}$=0.7651 days). Cleaned and interpolated Kepler light curves (top panels) and their short-term Fourier-transforms at the double rotation frequency (middle panels) and at the rotation frequency itself (bottom panels). Modulations indicate roughly 330(50) days and 410(50) days long activity cycles for KIC 03541346 and KIC 05791720, respectively.*

## 2.3.4 Cycles and the dynamo

The basic goal to derive activity cycles is to study the behavior of the underlying dynamo. Baliunas et al. (1996) revealed the first time the relationship between the cycle length and the rotational period which is proportional to the dynamo number, i.e., $P_{cyc}/P_{rot} \sim D^\iota$, where $\iota \geq 1/3$. Later, Oláh et al. (2009) refined this relation using long-term datasets of 20 active stars of different types (dwarfs, giants, singles and binaries). Recently, a number of cycles have been derived for fast-rotating K and M dwarf stars with rotational periods less than 1 day, which populate the short period end of the relation (Vida et al. 2014). Except for M-dwarfs, $\iota$ is about 0.8, showing that longer period stars have longer cycles. For the whole sample of M-dwarfs the slope of the fit between $\log(P_{cyc}/P_{rot})$ and $\log(1/P_{rot})$ is about unity (see Fig. 10, and cf. Savanov 2012), i.e., the cycle lengths for M-dwarfs do not depend on the rotational period. This fact implies that the magnetic activity of M-dwarfs is driven by a different kind of dynamo (likely $\alpha^2$), while the majority of the G-K stars (dwarfs and giants) are thought to harbor $\alpha\Omega$ type dynamos.

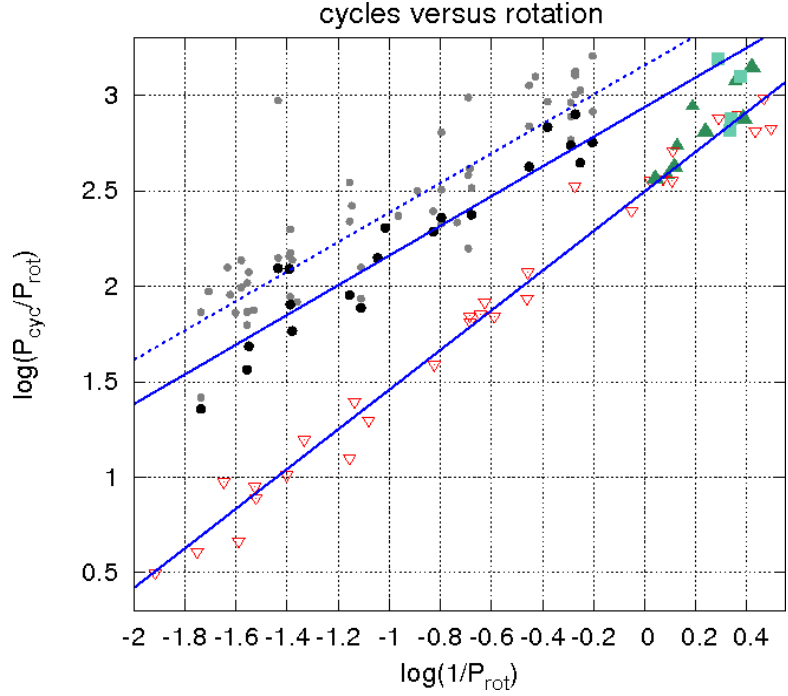

*Fig. 10: Relationship between rotation and cycle lengths of active stars of different types. Black dots stand for the shortest cycle lengths from the multiple cycles of the same star, grey dots show the longer cycles from Oláh et al. (2009). Filled squares and triangles represent K-M dwarf stars from Vida et al. (2013) and Vida et al. (2014). Dotted line is the fit for all of the cycles, while the upper solid line is the fit for the shortest cycles only, excluding all the M-dwarfs in both fits. Open triangles are cycles derived for the M-dwarf sample by Savanov (2012). Bottom line is the fit to Savanov's (2012) sample. See the text for further explanation.*

Some attempts have already been made to model stellar dynamos with oscillating nature, based on a solar-type model. Işik et al. (2011) used an $\alpha\Omega$ type dynamo following the rise of flux tubes through the convection zone, which resulted in a consistent determination of the emergence latitudes and tilt angles, and added horizontal flux transport at the surface. With the appropriate changes of the input model parameters butterfly diagrams with their cycles could be determined for rapidly rotating dwarf and subgiant stars, as well as for the Sun. It is worth to mention that a model dynamo with the parameters of the K1-subgiant V711 Tau ($P_{rot}$ = 2.8 days) shows well-defined cycles (Işik et al. 2011, Fig. 12). Calculating the model for a time base equal to the observations of V711 Tau, double cycles are found on timescales of 4-5 years and 10-20 years, which were similar to the results from observations. Although the model gives the total unsigned magnetic flux and not the brightness, the magnetic flux should be strongly correlated with the activity level, thus the brightness of the star (Işik 2012). Important factor is, that concerning the modeling, the shorter cycle originates simply from the dynamo action in the stellar interior while the long term cycle-like feature is due to stochastic flux emergence in the model. This result may shed some light on the very important question: from the *observed* multiple cycles which one comes from the dynamo action and which one(s) is (are) due to other effects.

Just recently, from modeling activity cycle lengths Dubé & Charbonneau (2013) have found a relation between the rotational and cycle periods, showing that the existence of such a connection is quite robust. However, their relation showed that stars with longer rotational periods had shorter cycles, in contradiction with quite a few results (e.g., Oláh et al. 2009, Vida et al. 2014)

## 3 Observing the dynamo by surface reconstruction

Imaging of stellar surface structures is a commonly used technique to observe ingredients of the underlying magnetic dynamo. Sunspots are deeply anchored into the surrounding plasma, they follow the large scale plasma flows due to confinement, thus providing diagnostic marks on the surface differential rotation (Paternò 2010) and maybe other large scale velocity fields. We can see sunspots following also local flow fields such as, e.g., convective flows, the geostrophic flow (Spruit 2003) or performing asymmetric proper motions due the tilt of the emerging flux (van Driel-Gesztelyi 1997), which, on the other hand, can totally overwhelm flows on larger scales. That is why the surface flow pattern of the solar meridional circulation (Zhao et al. 2013) cannot be observed easily from tracking sunspots (Wöhl 2002, but cf. e.g., with Komm et al. 1993).

We expect that positions, extensions and motions of starspots on cool stars will also provide information on the characteristics of the underlying dynamo. So, for want of better, imaging and, if possible, tracing stellar surface structures on magnetically active stars are essential tools to observe stellar dynamo ingredients. However, compared to the Sun, direct imaging of the stellar surfaces are not possible. Thus, reconstructions should rely on indirect inversions, such like Doppler imaging and Zeeman-Doppler imaging (see Sect. 3.1.2). But the time resolution of such inversions is necessarily in the order of the rotation period of the star and surface variability on a much shorter term will just increase the noise. Nevertheless, during the past decades the observing facilities and the inversion techniques underwent a significant development and stellar surfaces can be studied in even more detail.

In this section we overview the most advanced techniques for magnetic activity research, with flashing some of the most important results obtained with them so far.

## *3.1 Inversion techniques*

### 3.1.1 Photometric modeling

Since the mid 70's different photometric spot models have been developed to study starspot-distorted light curves of active stars. The observed rotational variability has been modeled from 1-D photometric information by spot modeling techniques of different approaches. The basic idea of such techniques is to find a geometrical model (i.e., surface spot distribution) to explain the rotational modulation of the observed light curve. These techniques, either analytic or numeric, apply various fitting algorithms to iteratively minimize the residuals between the observed and the theoretical light curves (see e.g. Budding 1977, Kang & Wilson 1989, Eker 1994, etc.).

With the considerable computational advances from the 1990's, such innovations came to the front like the time-series spot model using the photometric time series to derive a consistently evolving spot model (e.g., Strassmeier & Bopp 1992, Oláh et al. 1997, Ribárik et al. 2003); models assuming spots with two temperature components, i.e. spots with umbral and penumbral parts as seen on the Sun (Amado et al. 2000); and models applying randomized multispot solutions (Eaton et al. 1996). This latter provided a new scope of involving differential rotation, as well as handling the inconsistency between numerous smaller surface structures recovered by Doppler imaging (Sect. 3.2.1) and the traditional two- or three-spot solutions. However, due to the dramatic rise of free parameters, results of such multispot solutions could only be assessed statistically.

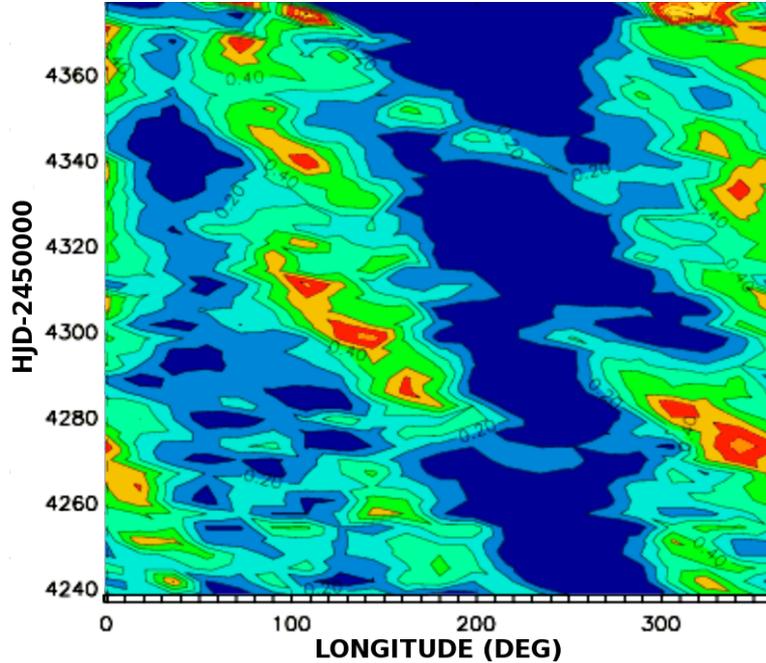

*Fig. 11: Light curve inversions of the spotted surface of CoRoT-2a from Lanza et al. (2009). The plot shows the distribution of the spot covering factor f vs. longitude and time. Yellowish regions indicate maximum spot filling factor, dark blue regions indicate minimum coverage. Note the active regions forming, developing and fading away, and after their formation they perform a retrograde migration.*

The stability of the different spot solutions has been a real problem since the advent of these kinds of techniques. In so far, the information content of a 1-D time series is obviously limited, and using photometry alone for spot modeling is an ill-posed problem. Kővári & Bartus (1997) have demonstrated that virtually, a simplified two-spot model could provide sufficiently good fit for more complex multiple spot configurations. However, even if having additional spatial information on the spot distribution (for instance, when the inclination of the rotation axis is known, or the spotted star is a component of an eclipsing binary system and spot eclipses occur), the uniqueness of the recovered spot solution is seriously restricted by the data quality. Studies concerning this issue have been carried out by e.g. Vogt (1981), Strassmeier (1988), Rhodes et. al (1990), Eker (1995), Kővári & Bartus (1997), etc.

To overcome the uniqueness and stability problems, Lanza et al. (1998) introduced a regularizing function into the inversion process. According to their spot model, the surface map is built up from a given set of small surface elements and the surface intensity of every pixel is a function of its spot filling factor ($f_i$). As *a priori* assumption, the surface elements have to fulfill either the maximum entropy criterion (e.g., Nityananda & Narayan 1982) or the Tikhonov criterion (Tikhonov & Goncharsky 1987). The application of such a regularization proved to be very useful in modeling light curves (e.g., Roettenbacher et al. 2013), as well as in spectroscopic Doppler imaging techniques.

Just recently, Walkowicz et al. (2013) performed extended numerical experiments to assess degeneracies in models of spotted light curves. They confirmed that in the absence of additional constraints on the stellar inclination (e.g., *v* sin *i* measurements or occultations of starspots by planetary transits) spot latitudes could not be determined, not even when having ultra-high precision (i.e. *Kepler*) photometry. According to their experience, from spot modeling of stars with different rotation rates, subtle signatures of differential rotation can be measured, which may provide information on the distribution of spots; see Fig. 11 for an example. Also, the outstanding quality of space-photometry from *MOST*, *CoRoT* and *Kepler* missions compelled the development of new data processing techniques (e.g., in Lanza et al. 2009, Pagano et al. 2009, Bonomo & Lanza 2012, Herrero et al. 2013, García et al. 2013, Reinhold & Reiners 2013, etc.), since the available methods were not suitable anymore. See also Savanov's (2013) excellent review in this subject.

### 3.1.2 Doppler imaging, Zeeman-Doppler imaging

The basics of mapping nonuniform chemical abundances of a stellar surface from spectral lines were formulated more than half a century ago by Deutsch (1958). Later, model calculations (Khokhlova 1976) demonstrated the capability of investigating the spotted surface of Ap stars from a set of spectral line profiles. Goncharskii et al. (1977, 1982) proposed mathematical solution for the inverse problem of surface mapping by reconstruction of local line profiles from the observational data. The term "Doppler imaging" was initiated by Steven Vogt and his colleagues (Vogt & Penrod 1983) for their technique to map the active regions of late-type stars, when the Doppler broadening due to rotation surpasses all other broadening mechanisms. When applying a cool star spot on the surface, a nub will appear on the broadened photospheric line profile just at the wavelength that corresponds to the apparent (Doppler-shifted) radial velocity of the spot location at the disk. Namely, a high-resolution spectral line can be regarded as a 1-D snapshot of the 2-D surface. Thus, a suitable set of spectra collected at different rotational phases can be assembled into a reconstructed surface image. Vogt et al. (1987) demonstrated through tests that their improved Doppler imaging technique was able to readily recover surface structures of 15° angular size. For today, with the application of the most advanced observational and computing techniques, this theoretical resolution has increased up to 5°, i.e., the size of a large sunspot.

Beside the success in applying Doppler imaging for an ever growing number of stars one should not disregard the limitations of its scope. First of all, the rotational broadening should be dominant among possible line broadening mechanisms. Moreover, the projected equatorial rotational velocity $v \sin i$ should exceed 20 km s$^{-1}$. However, above a certain limit around $v \sin i = 100$ km s$^{-1}$ the profiles become shallow and the line distortions due to spots (typically 1% of the continuum) cannot easily be observed (cf. Vogt 1988). Among cool stars, principally, two groups fulfill this criterion: fast rotator PMS and MS dwarfs with $P_{rot}$ of the order of a few days, and moderately rotating ($P_{rot}$ around 10-20 days) subgiants or giants, which are often members of RS CVn-type close binary systems, that help maintain relatively fast rotation by synchronization. Due to the strong relation between rotation and magnetic field generation, its manifestations in dark spots are expected to be less characteristic for slower rotators. Note though, that surfaces of cool MS stars rotating at the solar rate cannot be resolved by the Doppler technique due to the inefficient rotational broadening.

Magnetic surface structures can be investigated by Zeeman-Doppler imaging (hereafter ZDI, Semel 1989, Donati et al. 1989), which works in a similar way as described before, but uses polarization signals of the Stokes profiles to recover the magnetic field geometry. Originally, ZDI was based on the weak field approximation, i.e., the circular polarization signal by the Zeeman effect (Stokes V) is assumed to be proportional to the first derivative of the unpolarized intensity (Stokes I) signal. But detecting Stokes parameters in atomic spectra is complicated, since the expected signatures are far below the noise level. Thence, signal-to-noise ratio of the polarization signals were generally boosted by the multi-line least squares deconvolution technique (e.g., Donati et al. 1997, see also Tkachenko et al. 2013 for a recent improvement). If all four Stokes signals are available, ideally, the surface magnetic field distribution can be reconstructed (Piskunov 1985). The linear polarization profile (Stokes Q and U) signatures, however, are even smaller and more obscured, therefore only Stokes V and I were used from the beginning (e.g., Brown et al. 1991, Donati & Brown 1997). In a sense, with this reduction ZDI was an *inherently undetermined* problem, which necessarily anticipated the question of reliability (e.g., Wehlau & Rice 1993, Berdyugina 2005, Rosén & Kochukov 2012, etc.). Attempts, however has already been made to overcome some of the difficulties regarding the interpretability, e.g., by restricting to certain field configurations (e.g., Donati & Brown 1997, Hussain et al. 1998, 2001, Piskunov & Kochukov 2002). In the last few years new horizons were opened up with high-resolution spectropolarimetric observations (HARPSPol, ESPaDOnS, Narval) and eventually the next-generation of magnetic Doppler imaging is prepared to be suited for full reconstruction of the surface magnetic topology (e.g., Rusomarov et al. 2013, Silvester et al. 2014).

Before long, the ultra-high resolution spectropolarimeter PEPSI@Large Binocular Telescope (Strassmeier et al. 2007) will provide the capability of observing all four Stokes signatures for a considerably large sample of late-type stars. For these upcoming data a new Zeeman-Doppler reconstruction technique is developed (Carroll et al. 2007, 2012); see Fig. 12. Through iterative regularization, this new ZDI code iMap solves the full polarized radiative transfer using an artificial neural network (Carroll et al. 2008). Indeed, iMap performs simultaneous reconstructions for the radial, azimuthal and meridional magnetic fields, as well as the surface temperature distribution map. This development provides further improvements regarding the overall scope and reliability of ZDI, since ignoring temperature inhomogeneities yields unreliable magnetic field reconstruction (Rosén & Kochukov 2012).

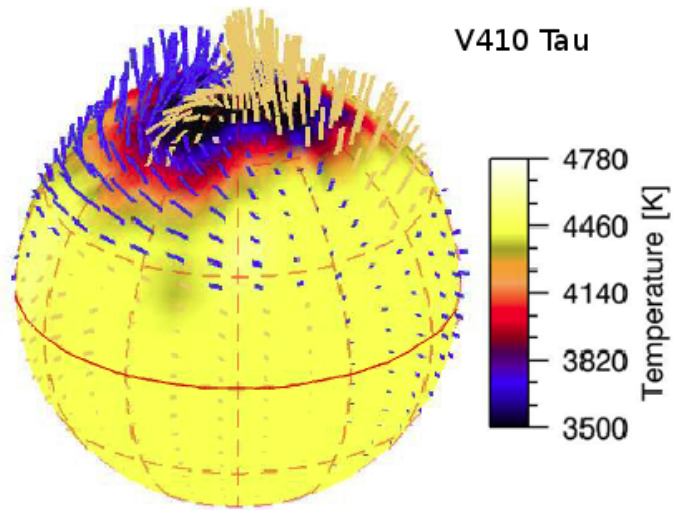

*Fig. 12: Zeeman Doppler image of the "young Sun" V410 Tau obtained with iMap (after Carroll et al. 2012) reveals an S-shaped twisted bipolar field geometry over the cool polar spot. Blue field lines are of negative polarity, while pale yellow lines are of positive. Vector lengths are proportional to the magnetic field strength of up to ±1.9 kG. Color background on the surface indicates temperature distribution according to the colorbar.*

## *3.2 Basic results from photometry*

### 3.2.1 Active longitudes

Solar observations revealed that longitudinal concentrations of various magnetic activity indicators existed from sunspots through chromospheric magnetic fields and flares, extending to coronal streamers and heliospheric fields. Statistical analyses of this kind of phenomena were carried out by several authors (e.g., Bumba & Howard 1969, Bumba et al. 2000, Berdyugina & Usoskin 2003, Kitchatinov & Olemskoy 2005, Li 2011, etc.). Albeit having important consequences for the dynamo regime, the grounds of the formation of such non-axisymmetry of solar activity is not clearly understood; but see, e.g., Moss et al. (1995), Ruzmaikin (1998), Weber et al. (2013), etc., for some of the possible explanations.

Long-lived active longitudes are known also from stellar observations (see, e.g., Oláh et al. 1991, Berdyugina & Tuominen 1998, Rodonò et al. 2000, Korhonen et al. 2002, Järvinen et al. 2005, Lanza et al. 2009, Lehtinen et al. 2011, etc.). Often, two permanent activity centers are detected, being separated longitudinally by half of the rotation phase on average (see also the "flip-flop" phenomenon in Sect. 3.2.2). Active stars in RS CVn-type close binary systems often show active longitudes that are locked in the frame of the revolving system (Oláh 2006). In these cases tidal forces and/or magnetic interaction are expected to play a significant role in locking the activity centers at preferred longitudes, which are often quadratures (Oláh 2006, Kővári et al. 2007a, Korhonen et al. 2010) or at the substellar point of the system and opposite to it (Oláh 2006 and references therein).

From the point of view of the mean-field dynamo theory, the existence of active longitudes requires breaking of the axial symmetry. Ruzmaikin (1998) suggested that non-axisymmetric mean field at the base of the convection zone could produce clustering of activity at certain longitudes. In the rapid rotation regime non-axisymmetric non-stationary mode can be excited (Moss et al. 1995, Moss 2004). Indeed, active longitudes are expected to occur quite widely (Moss 2005). On the other hand, in this regime the role of differential rotation is expected to be less significant. For close and contact binaries an extension of the mean field dynamo model is developed (Moss & Tuominen 1997) which can explain some of the observed features reviewed e.g. in Oláh (2006). New results for purely $\alpha^2$ type oscillatory solution have already been presented (Mantere et al. 2013). Recently, employing a thin flux tube model subject to a turbulent solar-like convective velocity field Weber et al. (2013) have found that large-scale flux emergence patterns exhibit active longitude-like behavior.

The appearance of active longitudes in close binary systems was modeled in detail by Holzwarth & Schüssler (2003). Their main conclusion was, that the erupting flux tubes were considerably affected by the companion star, depending on the tidal forces and the non-sphericity of the active star. The depth of the convection zone, the initial latitude of the flux tube in the bottom of the convection zone and the strength of the magnetic field together determine the place of the eruption. Details of all parameters are thoroughly discussed in Holzwarth & Schüssler (2003), comparing the models with some observational results.

## 3.2.2 Flip-flops

Jetsu et al. (1993) studied the active longitudes of the rapidly rotating giant FK Comae using 25 years of photometric data and found that the spot activity concentrated around two active longitudes which were separated by half of the rotation phase in longitude. One of them was found more dominant for a given period, and from time to time the dominancy switched between them. This phenomenon was named as the "flip-flop mechanism" (Jetsu et al. 1991, 1993) and was found to occur quite commonly among late-type stars showing spot activity.

Recent developments successfully introduced flip-flops into dynamo theory. According to Berdyugina et al. (2002) the coexistence of oscillating axisymmetric and stationary non-axisymmetric modes can explain active longitudes and flip-flop cycles on the solar-type K2-dwarf LQ Hya. Fluri & Berdyugina (2004) suggested a new approach to interpret flip-flops in terms of different combinations of the dynamo modes, enabling a fast comparison of computed solutions with observations (see also Moss 2004, Fluri & Berdyugina 2005). Flip-flop phenomenon was recovered from $\alpha^2\Omega$ type dynamo for weakly differentially rotating stars (Korhonen & Elstner 2005, Elstner & Korhonen 2005).

Observing flip-flop phenomenon, however, still remains challenging. First of all, because enough long and continuous time series are scarcely available. Moreover, phase drifts of the active longitudes are often recorded (e.g. Korhonen et al. 2004), thus, the interpretation of the observed phase jumps can be quite difficult, sometimes confusing or arbitrary, cf. Fig. 13, and see also Korhonen et al. (2004, Fig. 3) or Oláh et al. (2013, Fig. 8). Extensive Doppler imaging studies of this phenomenon would also be useful, but until now, only a few such results have been obtained (Vogt et al. 1998, Korhonen et al. 2004, Washuettl et al. 2009). The observational evidence proving that flip-flops occur together with polarity change between the two active longitudes is still to be found. Furthermore, the relation between spot cycles and flip-flop cycles is unclear and flip-flops can occur more often than the expected solar-like spot cycle (cf. Berdyugina 2005, and references therein), i.e., there is no *standard* representation of the observed flip-flop phenomenon.

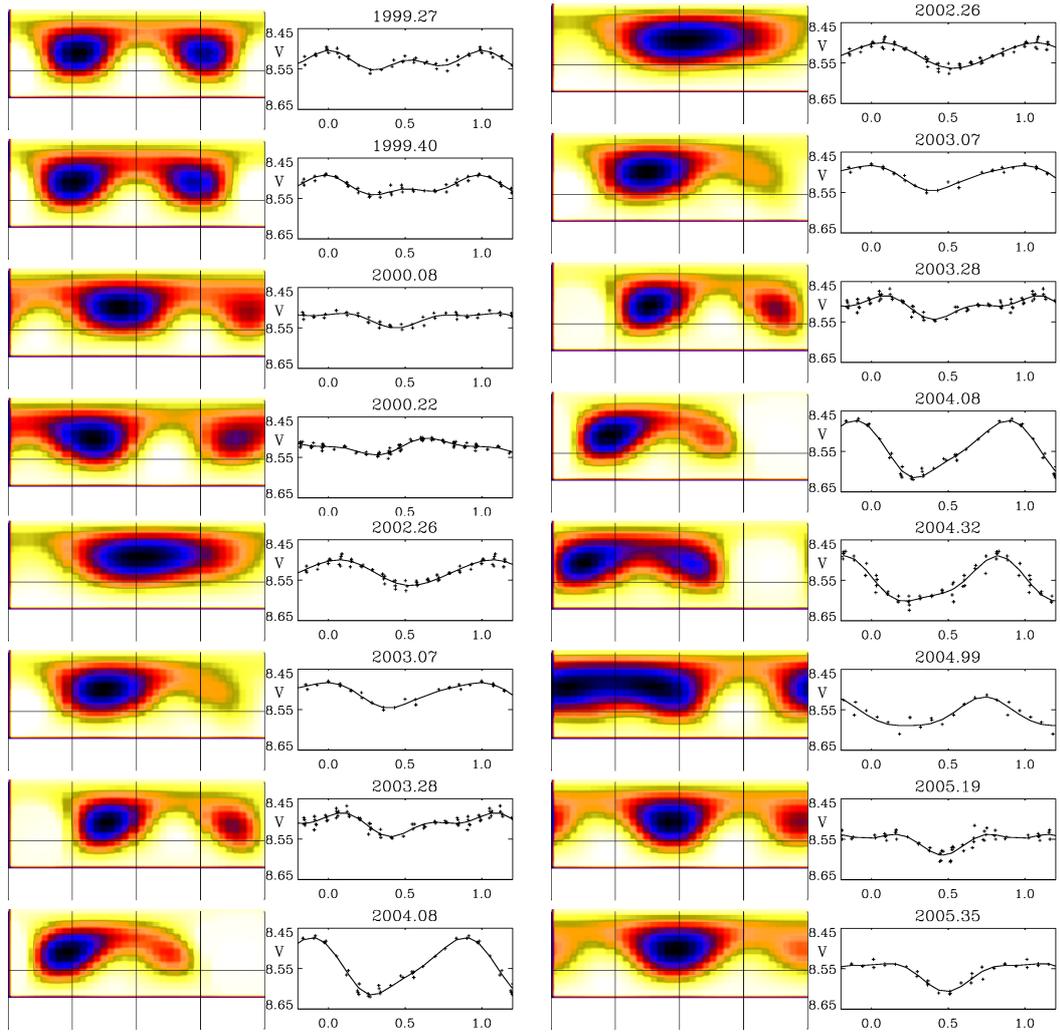

*Fig. 13: Light curve inversions for DP CVn from Kővári et al. (2013). Together with the surface maps, in the right panels plotted are the observations (crosses) and calculated V and y light curves (solid lines) for different seasons during 1999-2005. Derived maps usually show two active regions, separated by roughly half of the rotation phase. This pattern may bear some evidence of the flip-flop phenomenon, however, phase drifts and jumps would allow only an ambiguous interpretation.*

### 3.2.3 Differential rotation

Solar surface differential rotation is one of the most prominent manifestations of the magnetic dynamo working underneath. The phenomenon that sunspots close to the equator travelled across the disk faster than those located at higher latitudes was noticed first by Christoph Scheiner in the 17$^{th}$ century, just at the birth of modern observational astronomy. Centuries later, from the Greenwich sunspot records Maunder & Maunder (1905) derived a differential rotation coefficient $\alpha$ =

($P_{eq}$−$P_{pole}$)/$P_{eq}$ of 0.19 (note that conventionally, solar-type differential rotation is characterized by $\alpha$ of positive sign, while antisolar type differential rotation, when equator rotates most slowly, is featured by a negative shear parameter). More precise rotation law was derived from sunspot motions e.g., by Ward (1966). Spectroscopic measurements, however, yielded slightly different results for the photospheric material (see the review by Paternò 2010).

The surface rotation pattern imply firm constraints on the underlying dynamo process. However, the detection of surface differential rotation on cool stars is still a challenging observational task. Direct tracing of star spots is a potential way to observe surface shear pattern, however, it requires reliable surface reconstruction, such as Doppler imaging (see Sect. 3.2.4). Still, on a differentially rotating surface, a latitudinally drifting spot would produce photometric period change, therefore, seasonal changes of the rotation period ($\Delta P/P_{phot}$) derived from long-term photometric datasets can also be used as a clue for the shear parameter. Due to the lack of spatial resolution of spot distributions, light curve analysis enables only estimating the magnitude of the surface shear without sign. However, a transiting companion can help overcome this flaw. With the advantages of the eclipse mapping technique Huber et al. (2009, 2010) obtained information on the fine structure of the spot distribution of CoRoT-2a (see also Silva-Valio & Lanza 2011, Nutzman et al. 2011). Roettenbacher et al. (2011) studied the differential rotation of II Peg by inverting a set of light curves into surface maps, which were used to infer the surface evolution.

Just recently, Reinhold & Reiners (2013) developed a fast method for determining differential rotation for spotted stars by searching for different (but similar) periods in their light curves presumably caused by spots rotating at different rates. This method is suitable for large datasets, such like the *Kepler* database, and is capable of measuring rotation periods and detecting differential rotation as well, within statistically reasonable errors, however, without distinction between solar and antisolar type rotation profiles. The shear parameter is estimated as $\alpha := |P_2-P_1|/\max\{P_1,P_2\}$, where $P_1$ and $P_2$ are the primary and secondary periods, respectively. According to the upcoming application for thousands of *Kepler* stars in Reinhold et al. (2013), the shear coefficient grows towards longer periods and slightly increases towards lower temperatures, supporting previous findings from either observations (Barnes et al. 2005) or theory (Küker & Rüdiger 2011).

The possibility of measuring differential rotation for solar-type stars through asteroseismology was investigated first by Gizon & Solanki (2004). The feasibility of detecting differential rotation by this technique depends strongly on the precision of the frequency measurements, as well as on the stellar inclination. Moreover, further difficulties can be introduced by magnetic fields, large scale surface flow fields, etc. (see Lund et al. 2014 and the references therein). On the other hand, Lund et al. (2014) also showed, that frequency splittings could indeed be used in the future to test whether the latitudinal differential rotation is solar-like or antisolar-like (see Sect. 3.3.3), even when having not very precise measurements of frequency splittings.

## *3.3 Surface flows from spatially resolved surface maps*

The inherent ability of studying stellar surfaces by the means of Doppler imaging was considerably extended with employing time-series spectroscopic datasets covering two or more subsequent rotation periods (cf., e.g., Donati & Collier Cameron 1997, Strassmeier & Bartus 2000, Kővári et al. 2004, 2007a, 2007b, 2012, etc.). Time-series Doppler imaging has been proved to be extremely useful in observing stellar dynamo ingredients. With the use of subsequent surface maps such features can be investigated as the spot lifetime and evolution, or the structural change of the spot distribution due to surface differential rotation, etc. However, again, not only high-resolution good quality data but also reliable reconstructions and thoroughly tested data processing techniques are needed. In the next sections we give a brief overview of the basic techniques and highlight some of the recent results of this field.

### 3.3.1 Spot tracking, cross-correlation, sheared image

Tracking of surface features on the Sun has widely been used to infer the time evolution of solar surface structures from time series of images. In the same way, time-series Doppler images may reveal information on short-term spot changes and help study dynamo actions in stellar analogs. Longitudinal migration of long-lived spots at different latitudes were compared to derive the latitude dependent rotation law for the active K1-subgiant component of V711 Tau (=HR 1099, Vogt et al. 1999), i.e., one of the most studied RS CVn-type binary systems (see e.g., Donati 1999, Strassmeier & Bartus 2000, Ayres et al. 2001, Donati et al. 2003, Petit et al. 2004, Berdyugina & Henry 2007, etc.). Authors found a weak ($\alpha = -0.0035$) differential rotation of antisolar type. Pole was found not only to rotate faster than the equator, but to be nearly synchronized with the orbit, suggesting a high degree of tidal coupling (cf. Scharlemann 1982).

A natural possibility is to use Doppler imaging for deriving cycle lengths as well, since this method gives a direct measure of the total area of spots on the stellar surface. It needs, though, systematic, long-term spectroscopic observations of high resolution and with good phase coverage for each image. From 11-years of Doppler images Vogt et al. (1999), found an about 3 years long cycle (±0.2 yr) from the area of both the polar and low latitude spots of V711 Tau. This value is very close to the shortest cycle of 3.3 years obtained for V711 Tau from long-term photometry (Oláh et al. 2009).

The identification of long-lived spots from one observing season to the next (typically from year to year), however, can be difficult, especially, when spot structures become unstable and change significantly on yearly timescale. For this kind of studies, therefore, subsequent Doppler images sampled close to each other proved to be more suitable. From those image pairs, differential rotation can be derived from the technique of cross-correlation instead of simple tracking of individual features. Actually, the very first measurement of surface differential rotation on a star was applied cross-correlation (Donati & Collier Cameron 1997). The technique is based on computing numerical cross-correlation functions along the longitude for all latitudinal slices of the maps to be compared (maps are usually cut into slices of 5° width, i.e., the resolution limit). The cross-correlation functions are then assembled into a 2-D cross-correlation function map which can instantly reveal the differential rotation pattern (see e.g., Weber & Strassmeier 1998, 2001, Skelly et al. 2008, Kővári et al. 2013, etc.). Nevertheless, even when having two subsequent surface maps with small time lag between them, rapid spot evolution (e.g., vivid interaction between emerging flux and its surroundings) is always of concern, since it can easily mask the differential rotation pattern. To boost, however, the signature of the surface shear in the correlation pattern Kővári et al. (2004, 2007a, 2013, 2014a) developed the technique 'ACCORD' (acronym from 'average cross-correlation of time-series Doppler images' ) for a contiguous set of Doppler reconstructions. The key advantage of the method is that averaging cross-correlation maps definitely emphasizes jointly present features (i.e., the surface shear pattern itself) while suppresses false detections in the correlation pattern by unwanted effects (e.g., from rapid spot evolution, crosstalk between neighboring spots, as well as spurious features from data noise). The capability and reliability of this method were demonstrated on a series of elaborated tests (Kővári et al. 2014b), see Fig. 14.

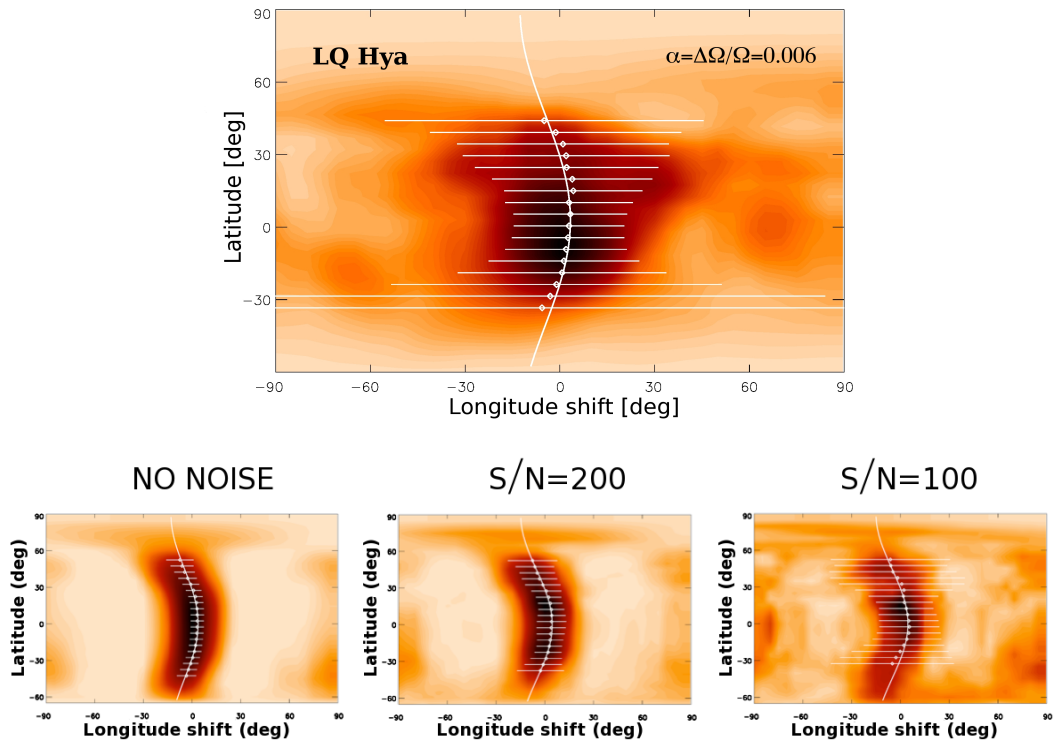

*Fig. 14: Top: observed differential rotation of the fast-rotating K2-dwarf LQ Hya resulting in a weak (α = 0.006) solar-type surface shear (Kővári et al. 2004). Bottom: Testing the reliability of the average cross-correlation technique ACCORD (Kővári et al. 2014b). The surface shear of LQ Hya was applied to a set of artificial Doppler images (artificial spectroscopic sampling patterned an actual observing run). Then, average cross-correlations were performed and surface shear parameters were derived for different data quality, yielding α of 0.0058±0.0004, 0.0067±0.0033 and 0.0068±0.0048 values for no noise (left), S/N=200 (middle) and S/N=100 (right), respectively. Note that uncertainty originates also from imperfect image reconstruction due to uneven or gappy (i.e., realistic) phase coverage of the observations.*

Among the tracer-type techniques, local correlation tracking technique (LCT) should not be missed, even if it became a favorable tool in solar physics (e.g., Sobotka et al. 2000, Švanda et al. 2006). Attempts however, were made to demonstrate the power of LCT also on stellar observations (Kővári et al. 2007c, Vida et al. 2007). The technique is based on the principle of the best match of two image frames that record the tracked features at two subsequent instants. The resulting vector map can essentially be regarded as the surface flow field over the time lag between the initial frames, see Fig. 15. This approach could provide an important clue in observing global and local surface flows, however, further tests are needed to confirm solidity.

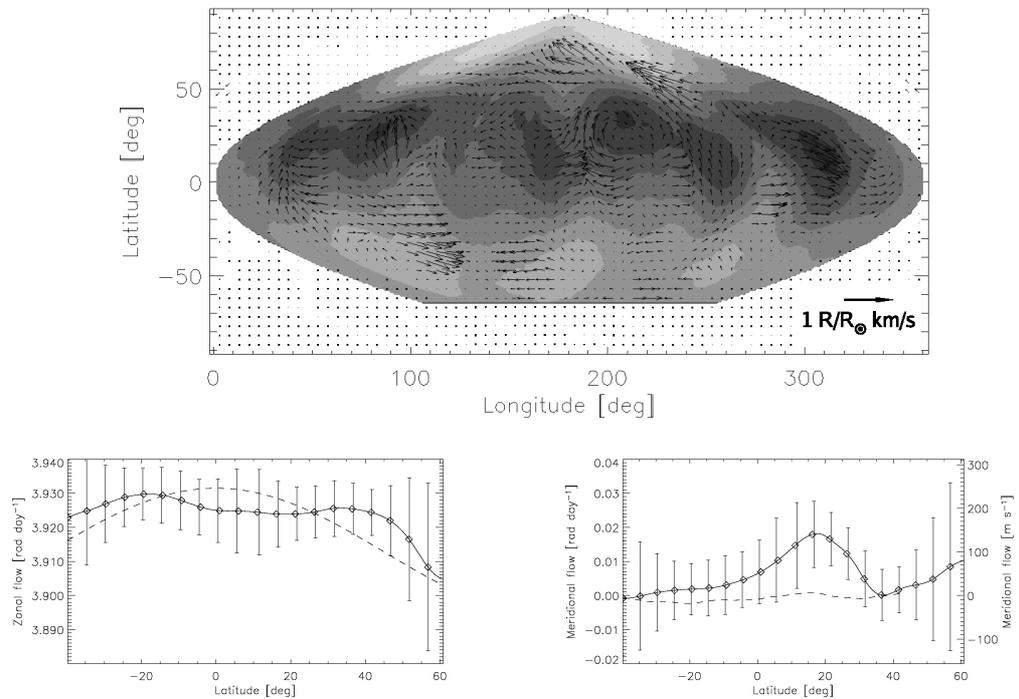

*Fig. 15: Top: Flows in the photosphere of LQ Hya by applying local correlation tracking (LCT) technique for a time series of 28 Doppler images. Bottom: Mean zonal flow component (left) suggests a weak solar-type differential rotation (dashed line), just consistently with the result in Kővári et al. (2004). Note, that the poleward flow implied by the mean meridional component (bottom right) is around the detection limit, therefore ambiguous.*

Surface differential rotation can be detected even if having data sufficient only for one single Doppler reconstruction. The parametric imaging method, often called sheared image method (e.g., Petit et al. 2002, Weber & Strassmeier 2005, Barnes et al 2005, Morgenthaler et al. 2012, etc.) involves surface shear as an additional parameter in the line profile inversion process. Computations are done for a range of meaningful shear parameters and the most likely value is chosen on the basis of goodness of fit. The disadvantage of this method, however, is that it introduces yet another free parameter into the inversion process, and a predefined latitudinal rotation law (generally a $\sin^2\beta$ type) is forced.

In some cases, however, the reliability of the results obtained from sheared image method is doubtful, as demonstrated by the following comparison. The differential rotation of the single early-G type V889 Her ($P_{\rm rot}$ = 1.337 days), a young Sun, was studied e.g., by Marsden et al. (2006) and Jeffers & Donati (2008) by the means of parametric Zeeman-Doppler imaging, and strong solar-type differential rotation was reported with $\alpha$ of 0.084 and ≈0.1, respectively. Note that such a large shear does not fit the empirical law by Barnes et al. (2005); see also Kitchatinov & Rüdiger (1995), Reiners (2006), Küker & Rüdiger (2011), Reinhold et al. (2013) on how the differential rotation is influenced by temperature. On the other hand, Järvinen et al. (2008) argued for a substantially weaker differential rotation, while Huber et al. (2009) preferred rigid rotation (still allowing a weak surface shear). A more recent parametric Doppler imaging study (Kővári et al. 2011) resulted in a weak solar-type rotation law with $\alpha = 0.009$. The numerical model by Kitchatinov & Olemskoy (2011) for a $T_{\rm eff}$=5800K dwarf rotating at the angular velocity of V889 Her would suggest $\alpha = 0.016$, i.e., much weaker than the values 0.084 and ≈0.1 found by Marsden et al. (2006) and Jeffers & Donati (2008), respectively. Such contradictory results might partially be explained by temporal variations on the differential rotation (cf. Donati et al. 2003), but Jeffers et al. (2010) found no evidence for it. Nevertheless, Kővári et al. (2014b) have demonstrated recently, that a large polar cap, such like the one found on V889 Her (see also Strassmeier et al. 2003a, Frasca et al. 2010), can distort the rotational profile similarly as the differential rotation does, consequently yielding false measure of the surface shear by parametric imaging technique; see Fig. 16.

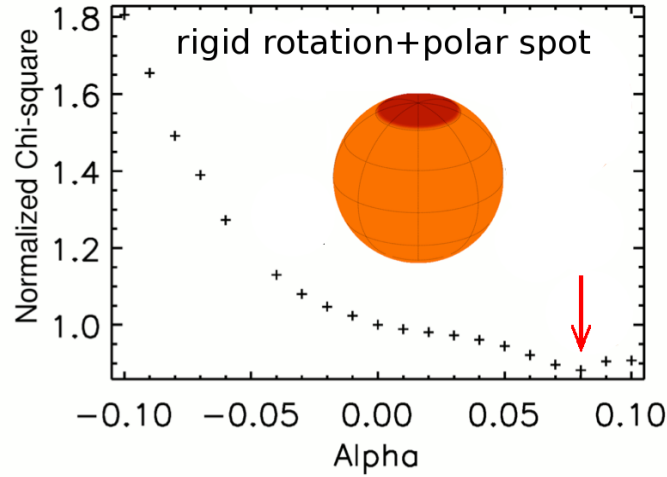

*Fig. 16: False detection of surface shear. Applying the parametric imaging study for an actually rigidly rotating test star with a large polar cap would result in solar-type differential rotation. Note the best fit at α = 0.08 instead of zero shear.*

### 3.3.2 Binarity and activity

In RS CVn-type active close binary systems tidal effects help maintain the fast rotation and so the magnetic activity at a higher level. In such an interaction, tidal coupling between the convective envelope of the evolved star and its companion is of utmost importance in understanding the dynamo action underneath (e.g., Scharlemann 1981, 1982, Schrijver & Zwaan 1991, etc.). Although earlier photometric techniques gave only a poor chance to study this interaction in detail, later on Doppler imaging and in some exceptional cases interferometric imaging (Monnier et al. 2007, Parks et al. 2012) represented observationally a great improvement. The behavior of the emerging magnetic flux tubes under the gravitational influence of a companion star was analyzed first by Holzwarth & Schüssler (2000, 2002). Recently, active components in different RS CVn-type systems were compared in Oláh et al. (2012) and in Kővári et al. (2012b) and marked differences were depicted regarding the spot distribution and the surface differential rotation.

One of the compared systems, V711 Tau (K1IV) showed very little evidence of differential rotation (cf. also Vogt et al. 1999) while from Zeeman Doppler imaging Petit et al. (2004) found a bit stronger shear, but still 40 times smaller than that of the Sun. The reduced differential rotation was attributed to the strong tidal interaction of the G5V companion star (see also Muneer et al. 2010). Congruently, the very similar system UX Ari (K0-K1IV) did not show sign of strong surface shear (Rosario et al. 2008), indeed, spot distribution was found to be fixed to the orbital frame for long (cf. Lanza et al. 2006). Contrarily, EI Eri (K1IV) showed period deviation $\Delta P/P_{orb}$ reaching ±2% (Oláh et al. 2012) and time-series Doppler imaging study yielded $\alpha$ = 0.037 differential rotation coefficient (Kővári et al. 2009), i.e., one-fifth of the solar shear. In accord with this, long-term Doppler imaging study by Washuettl et al. (2009) did not reflect any preferred (i.e., phase-locked) spot positions. As compared to the former cases of V711 Tau and UX Ari, the secondary component in EI Eri is relatively small (M4-5) and thus its gravitational influence on the dynamo hosting primary seems to be much less significant.

The two long-period RS CVn-type systems, σ Gem and ζ And have similarly K1-giant primary components (with $P_{orb}$ of 19.6 days and 17.8 days, respectively), both showing active longitudes locked mostly at quadrature positions in the binary reference frame (Kajatkari et al. 2014, Kővári et al. 2007a). However, gravitational distortions were found to be quite different (Kővári et al. 2012b and references therein), getting stronger deformation for ζ And, while either small or even negligible non-sphericity for σ Gem. The scaled graphs of the two systems, see Fig. 17 (see also Schrijver & Zwaan 1991, Fig. 1), however, reveal a basic difference in the mass distribution of the binary systems, which may account for some other observed differences. Surface differential rotations were determined and refined for both systems by Kővári et al. (2007a, 2007b, 2012a, 2012b) yielding solar-like rotation law with $\alpha$ = 0.055 shear parameter for ζ And and oppositely, antisolar type surface differential rotation with $\alpha$ = −0.04 for σ Gem. Using a unique cross-correlation technique (see also Sect. 3.3.4) applied to latitudinal displacements of surface features Kővári et al. (2007a, 2007b) investigated the meridional surface flows on ζ And and σ Gem. Interestingly, the results indicated quite different phenomena: a negligible or possibly a weak equatorward drift on ζ And, while a clear and much stronger poleward flow on σ Gem. This observation is indeed in agreement with theoretical expectations by Kitchatinov & Rüdiger (2004), see also Sect. 3.3.3, who attributed the antisolar differential rotation to strong meridional flow which could be caused e.g., by field forcing of a close companion. Such examples expose the importance of binarity in controlling the activity of the evolved components in close binaries e.g., by either driving or suppressing differential rotation, or by rewriting the subsurface scenario of the magnetic flux emergence (Holzwarth & Schüssler 2003).

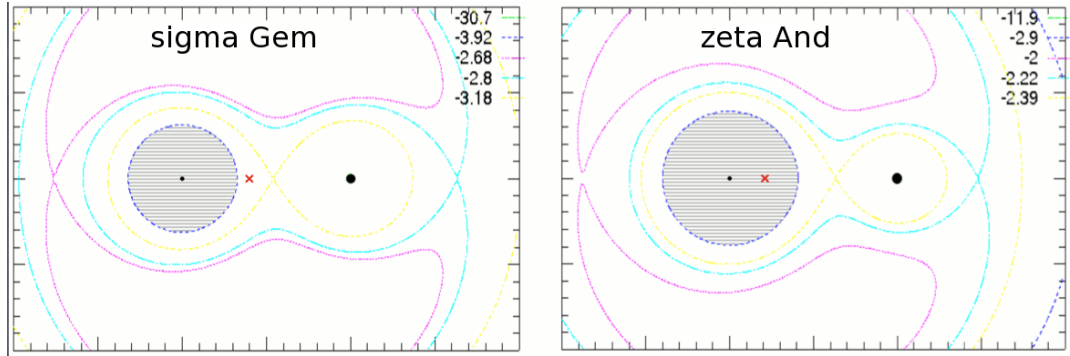

*Fig. 17: Roche potentials of σ Gem (left) and ζ And (right) from using* Nightfall (http://www.hs.uni-hamburg.de/DE/Ins/Per/Wichmann/Nightfall.html). *Cross-sections of the evolved components are grey shaded, gravity centers are marked with red crosses.*

### 3.3.3 Antisolar type differential rotation

Antisolar type differential rotation, when polar regions rotate faster than the equator, has been found for a small number of stars (e.g., Vogt & Hatzes 1991, Strassmeier et al. 2003b, Oláh et al. 2003, Weber et al. 2005, Weber 2007, Kővári et al. 2007b, 2013, Kriskovics et al. 2014, etc.). In Fig. 18 plotted are the derived surface shear coefficients vs. rotation period for a set of the most-observed targets of either solar or antisolar-type rotation profile. Regardless the sign, absolute values of the surface shear parameters seem to follow a weak dependence of the surface shear on $P_{rot}$ (cf. Küker & Rüdiger 2011). From the plot one can discerne that antisolar differential rotation was detected mainly for long-period giants being either single stars (e.g., DP CVn, DI Psc) or members in RS CVn-type binary systems (e.g., σ Gem, HK Lac).

To explain this formerly disregarded feature, first Kitchatinov & Rüdiger (2004) proposed a theoretical revision by considering a fast meridional flow. Differential rotation can be produced by meridional flow, which, on the other hand, can be fast when spherical symmetry is perturbed either gravitationally by a close companion or thermally by large cool spots. Simulations for a giant star (having 2.5 solar mass, 7.91 solar radius, $P_{rot} \approx 27$ days) showed that increasing poloidal field inferred the reversion of the meridional flow to poleward on the surface, and, at a certain rate, the differential rotation reversed to antisolar case (Kitchatinov & Rüdiger 2004, Kitchatinov 2006).

Further theoretical support for antisolar differential rotation has been provided in the recent studies by Chan (2010) and Gastine et al. (2014). The transition between the regimes of solar-type and antisolar differential rotation was investigated with 3-D simulations and it was found, that the direction of the rotational profile was dependent on the Coriolis force in the sense, that rapid rotators with large Coriolis numbers (or analogously, with small Rossby numbers) performed solar-like differential rotation, while antisolar type differential rotation was derived for moderate rotators having large Rossby numbers.

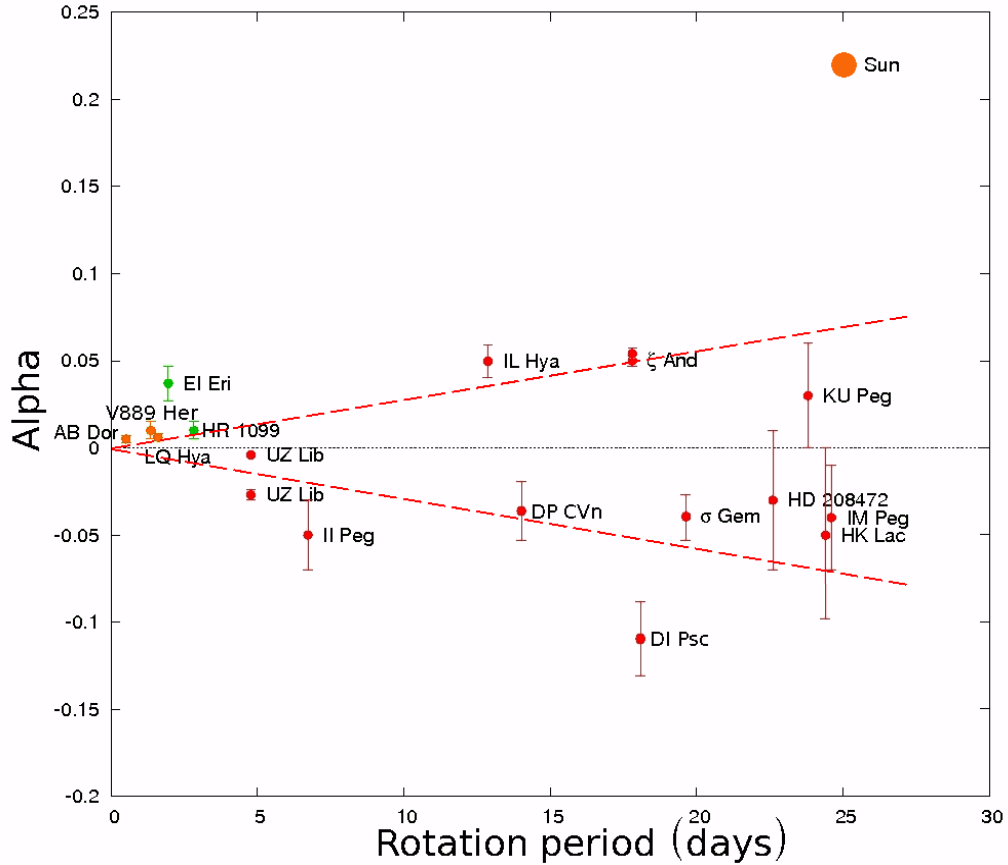

*Fig. 18: Differential rotation coefficients from Doppler imaging studies. The plot is an extended and updated version after Weber (2007). Absolute values of the surface shear parameters follow the empirical trend of Barnes et al. (2005) in the sense, that slower rotators perform larger surface shear. Dashed line (and its mirroring about the horizontal axis) is a simple linear fit to the absolute values of the shear parameters, giving a slope of ≈ 0.0028 $d^{-1}$.*

Recent investigations of two single giants, namely DP CVn and DI Psc (Kővári et al. 2013, Kriskovics et al. 2014, see also Fig. 18) drew the attention to a possible connection between antisolar differential rotation and unusually high surface lithium abundances observed in a handful of RGB giants, including DP CVn and DI Psc. The lithium is supposed to be transported from the inner parts onto the surface by an extra mixing mechanism (Charbonnel & Balachandran 2000), which, on the other hand, could also be responsible for transporting angular momentum towards the poles by meridional flows. This, eventually, would result in antisolar differential rotation, which was found for both targets. Moreover, antisolar differential rotation was found for another target of such kind, the high-Li K2-giant HD 31993 (Strassmeier et al. 2003b). We note, however, that further observations are needed to continue detections of differential rotation for similar objects and thus elaborate on this speculation.

### 3.3.4 Meridional flows: measurements and interpretations

Beside differential rotation, meridional circulation is the other key velocity pattern that is related to the dynamo. The influence of meridional circulation on solar and stellar dynamos was recognized and investigated by several authors (e.g., Choudhuri et al. 1995, Dikpati & Gilman 2001, Bonanno et al. 2002, 2006, Miesch 2005, Holzwarth et al. 2007, Jouve & Brun 2007, Pipin & Kosovichev 2011, Küker & Rüdiger 2011, etc.). Current helioseismic measurements suggested the existence of at least two meridional circulation cells in the Sun (Zhao et al. 2013). In turn, observing meridional flows on stars are much more difficult, since the meridional motion of surface features are expected to be at least one order of magnitude smaller compared to the zonal shifts at different latitudes due to surface differential rotation (note that the ratio of the average zonal and meridional surface flow velocities on the Sun is around 20:1). On the other hand, such tiny latitudinal motions could only be observed by employing reliable high resolution Doppler reconstructions. But note also, that, in general, stellar surface inversion techniques are less capable of recovering latitudinal information, especially, when the inclination angle cannot be determined precisely. Still, in particular cases, time-series Doppler imaging can be suitable to measure common meridional motion of surface features. The most reliable technique is similar to the one introduced in Sect. 3.3.1 for detecting differential rotation from average cross-correlation function maps, but contrary, cross-correlations are done along meridional circles.

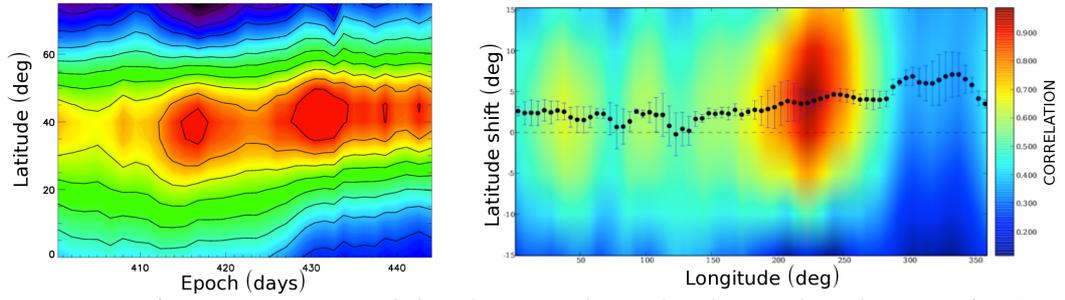

*Fig. 19: Left: synoptic map of the changing latitudinal spot distribution of σ Gem from 34 time-series Doppler images using the Ca I 6439 mapping line. An average poleward drift of the most prominent features during the covered term is striking, in accordance with the latitudinal cross-correlation function map (right). As a result, ≈220 ms$^{-1}$ flow velocity was derived (Kővári et al. in prep.).*

Due to observational difficulties, until now, some few determinations of meridional motions have been published (e.g., Strassmeier & Bartus 2000, Weber & Strassmeier 2001, Kővári et al. 2007a, 2007b, 2009, 2014a, in prep). The measured meridional flow velocities (of either equatorward or poleward) fall between $10^2$–$10^3$ ms$^{-1}$, i.e., stellar meridional flows are found to be 10–100 times faster than the solar meridional circulation on the surface (≈15 ms$^{-1}$). Note, though, that selection effect should also be considered, since flows of the solar magnitude cannot be observed on stars. A preliminary statistics (Weber 2007) indicated a weak dependence of the meridional flow on the dimensionless surface shear coefficient. The scatter, however, was too large to be suitable for deriving a proper relationship. Nevertheless, the trend in Weber (2007) would support the theoretical prediction in Kitchatinov & Rüdiger (2004), that strong meridional flow can drive antisolar differential rotation. One of the most spectacular results so far, supporting also this prediction, was obtained for σ Gem in Kővári et al. (2007a). Recently, Kővári et al. (in prep) have pointed out, that the method of latitudinal cross-correlation used in Kővári et al. (2007a) suffers from an incompleteness due to the singularity rising towards the pole. To avoid this imperfection, a refinement was carried out which yielded an average poleward drift of ≈220 ms$^{-1}$ on the surface of the K1-giant (see Fig. 19), i.e., a slightly smaller value compared to the result of ≈300 ms$^{-1}$ in Kővári et al. (2007a), but still in accordance with it.

## 4 Closing remarks and outlook

Through examples we gave an overview on the recent progress in observing dynamo action in cool stars. As for today, about $10^2$ active stars have been mapped by the means of Doppler imaging, i.e., still a poor sample for establishing general conclusions on stellar dynamos. Clearly, much larger sample of stars would be needed for answering such questions like what kind of dynamo mechanism operates in a given type (class, age, rotation, metallicity, binarity, etc.) of active star.

However, when using observational results derived by employing inversion techniques of different kinds one should not forget about the reliability issue. Most reliable detections of surface flows on stars, such as differential rotation and meridional circulation, would require good quality densely sampled homogeneous datasets, covering enough long time intervals for the best performance spatial and time resolution. So far, time-series spectroscopy has been proved to be an outstanding experience in studying stellar surface velocity fields. An exceptional possibility of collecting such long-term spectroscopic datasets for a small sample of active stars was the historical night-time program with the 1.5m McMath-Pierce telescope at Kitt Peak National Solar Observatory. The aim of extending this kind of observational experiences eventuated the 1.2-m mirror STELLA robotic telescopes at Teide Observatory, operated by the Leibniz-Institute for Astrophysics Potsdam (Weber et al. 2012). A step forward in studying surface phenomena of active stars would be a dedicated high precision spectroscopic instrument on an at least 4-m class monitoring telescope.

*Kepler* mission confirmed the necessity of ultra-high precision photometry also in case of investigating magnetic activity. A very important next step would be a space mission with similar high precision as of *Kepler*, but capable of multicolor photometry, this way gathering much more physical information on the observed objects. PLATO (Planetary Transits and Oscillations of stars) mission has already been selected as a part of ESA's Cosmic Vision 2015-25 Programme. By observing up to $10^6$ relatively nearby stars of different populations, PLATO's measurements will allow to calibrate the relationship between stellar age and rotation, as well as to study activity cycles. In addition, compared to CoRoT and *Kepler* targets, PLATO's nearby and so relatively bright stars of up to 4 mag will provide more opportunity for ground-based high-resolution spectroscopic follow-up observations (Rauher et al. 2013).

For observing stellar activity cycles at least the presently running few APTs should continue their missions. The time is running in one way: a missed observation cannot be replaced with a new one in such a scientific field which deals with continuously changing, developing entities, like stars. The timescales of activity cycles are in many cases longer than the human lifetime. The nearest example is the Gleissberg cycle of the Sun, which is something like 50-200 years long with changing length, and even the 11-year Schwabe cycle repeats only 4-5 times at most, during a career of an astronomer. Concerning our present knowledge, the timescales of the longer cycles from the multiple cycles of other stars are of similar lengths than the longer solar cycles. This fact should not be disregarded, indeed, it needs a kind of humble competence to invest to such projects both from scientists and fundings, that will serve not the present but the future generations of scientists.

From future observations using either ultra-high precision space photometry or high-resolution high signal-to-noise spectroscopy and spectropolarimetry or even just from continuation of long-term monitoring surveys by 1-m class telescopes, we expect a better understanding of dynamos working in cool stars and of course, in our Sun. Especially, we may get closer to understanding the role of magnetic activity in the formation and evolution of star-planet systems and eventually in the origin of life.

**Acknowledgements** ZsK is grateful for the invitation by the Convenors to the ISSI WS "The Solar Activity Cycle: Physical Causes and Consequences", held between 11-15 November 2013, in Bern, Schwitzerland. This work has been supported by the Hungarian Science Research Programs OTKA K-81421 and OTKA K-109276, the Lendület-2009 and Lendület-2012 Young Researchers' Programs of the Hungarian Academy of Sciences. The final publication is available at Springer via http://dx.doi.org/10.1007/s11214-014-0092-0.